\documentclass[twoside,12pt]{article}%
\usepackage{epsfig}
\usepackage{amsmath}
\usepackage{amsfonts}
\usepackage{amssymb}
\usepackage{graphicx}%
\newcommand{\be}{\begin{equation}}
\newcommand{\ee}{\end{equation}}
\newcommand{\bea}{\begin{eqnarray}}
\newcommand{\eea}{\end{eqnarray}}

\begin{document}

\title{ Pion-Photon TDAs in the NJL Model \thanks{This work has been supported by the 6th Framework Program
 of the European Commission No. 506078 and 
MEC (FPA 2007-65748-C02-01 and AP2005-5331)} }
\author{A. Courtoy $^{1}$ and S. Noguera $^{1}$\\$^{1} $ {\small Departamento de Fisica Teorica and Instituto de F\'{\i}sica
Corpuscular,} \\{\small Universidad de Valencia-CSIC, E-46100
Burjassot (Valencia), Spain.}} \maketitle

\begin{abstract}
The pion-photon Transition Distribution Amplitudes (TDAs) are
studied, treating the pion as a bound state in the sense of
Bethe-Salpeter, in the formalism of the NJL model. The
results obtained explicitly verify support, sum rules and polynomiality
conditions. The role of PCAC is highlighted.

\end{abstract}


Hard reactions provide important information for unveiling the
structure of hadrons. The large virtuality, $Q^{2}$, involved in
the processes allows the factorization of the hard (perturbative)
and soft (non-perturbative) contributions in their amplitudes. In
 recent years a large variety of processes governed by the
Generalized Parton Distributions (GPDs), like the Deeply Virtual
Compton Scattering, has been considered. A generalization of GPDs
to non-diagonal transitions has been proposed in
\cite{Pire:2004ie}. In particular, the easiest case to consider is
the pion-photon TDA, governing processes like
$\pi^{+}\pi^{-}\rightarrow\gamma^{\ast}\gamma$ or $\gamma
^{\ast}\pi^{+}\rightarrow\gamma\pi^{+}$ in the kinematical regime
where the virtual photon is highly virtual but with small momentum
transfer.\ At leading-twist, the vector and axial TDAs,
respectively $V(x,\xi,t)$ and $A(x,\xi,t)$, are defined as
\cite{Courtoy:2007vy} {\small
\begin{align}
\int\frac{dz^{-}}{2\pi}e^{ixP^{+}z^{-}}\left.  \left\langle
\gamma\left( p^{\prime}\right)  \right\vert \bar{q}\left(
-\frac{z}{2}\right)  \gamma ^{+}\hspace*{-0.05cm}\tau^{-}q\left(
\frac{z}{2}\right)  \left\vert \pi^+\left(  p\right)
\right\rangle \right\vert _{z^{+}=z^{\bot}=0}  &
=i\,e\,\varepsilon_{\nu}\,\epsilon^{+\nu\rho\sigma}\,P_{\rho}\,\Delta_{\sigma
}\,\frac{V^{\pi^{+}}\left(  x,\xi,t\right)  }{\sqrt{2}f_{\pi}}%
~,\label{vectcurr}\\
\int\frac{dz^{-}}{2\pi}e^{ixP^{+}z^{-}}\left.  \left\langle
\gamma\left( p^{\prime}\right)  \right\vert \bar{q}\left(
-\frac{z}{2}\right)  \gamma
^{+}\hspace*{-0.05cm}\gamma_{5}\tau^{-}q\left(  \frac{z}{2}\right)
\left\vert \pi^+\left(  p\right)  \right\rangle \right\vert
_{z^{+}=z^{\bot}=0}  & =e\,\left(
\vec{\varepsilon}^{\bot}\cdot\vec{\Delta}^{\bot}\right)
\frac{A^{\pi^{+}}\left(  x,\xi,t\right)  }{\sqrt{2}f_{\pi}}\nonumber\\
&  +e\,\left(  \varepsilon\cdot\Delta\right)
\frac{2\sqrt{2}f_{\pi}}{m_{\pi }^{2}-t}~\epsilon\left(  \xi\right)
~\phi\left(  \frac{x+\xi}{2\xi}\right)
\quad, \label{axcurr}%
\end{align}
}where $t=\Delta^{2}=(p^{\prime}-p)^{2}$, $P=\left(
p+p^{\prime}\right)  /2$, $\xi=\left(  p-p^{\prime}\right)
^{+}/2P^{+}$, $\epsilon\left(  \xi\right) =1$ for $\xi>0$ and $-1$
for $\xi<0$ and where $f_{\pi}=93$ MeV. For any four-vector
$v^{\mu},$ we have the light-cone coordinates $v^{\pm}=\left(
v^{0}\pm v^{3}\right)  /\sqrt{2}$ and the transverse components $\vec{v}%
^{\bot}=\left(  v^{1},v^{2}\right)  .$ Finally, $\phi\left(
x\right)  $ is the pion distribution amplitude (PDA).

Apart from the axial TDA $A(x,\xi,t)$, the axial current
Eq.(\ref{axcurr}) contains a pion pole contribution, which can be
understood as a consequence of PCAC because the axial current must
be coupled to the pion. This second term has been isolated in a
model independent way. Therefore, all the structure of the
incoming pion remains in $A\left(  x,\xi,t\right)  $. The pion
pole term is not a peculiarity of the pion-photon TDAs: a similar
contribution would be present in the Lorentz decomposition, in
terms of distribution amplitudes, of the axial current for any
pair of external particles. This term is only non-vanishing in the
ERBL region, i.e. the $x\in\left[  -\xi,\xi\right]  $ region,
whose kinematics allow the emission or absorption of a pion from
the initial state, which is described through the PDA.

The $\pi$-$\gamma$ TDAs are related to the vector and axial
transition form
factors through the sum rules%
\begin{equation}
\int_{-1}^{1}dx~V^{\pi^{+}}\left(  x,\xi,t\right)  =\frac{\sqrt{2}f_{\pi}%
}{m_{\pi}}F_{V}\left(  t\right)
~,~~~~~\int_{-1}^{1}dx~A^{\pi^{+}}\left( x,\xi,t\right)
=\frac{\sqrt{2}f_{\pi}}{m_{\pi}}F_{A}\left(  t\right)  \quad.
\label{sumerule}%
\end{equation}

As usual, we consider that the currents present in Eqs.
(\ref{vectcurr}) and (\ref{axcurr}) are dominated by the handbag
diagram. The method of calculation developed in
\cite{Theussl:2002xp} is here applied. The pion is treated as a
bound-state in a fully covariant manner using the Bethe-Salpeter
equation and solving it in the NJL model. Gauge invariance is
ensured by using the Pauli-Villars regularization scheme. All the
invariances of the problem are then preserved. As a consequence,
the correct support is obtained, i.e. $x\in\lbrack-1,1],$ vector
and axial TDAs obey the sum rules, Eq.(\ref{sumerule}), and the
polynomiality expansion is recovered in both cases. Moreover, for
the DGLAP region, we have obtained the isospin relations
\begin{equation}
V\left(  -x,\xi,t\right)  =-2V\left(  x,\xi,t\right)  ,~~~~A\left(
-x,\xi,t\right)  =2A\left(  x,\xi,t\right)  ,~~~~~~~~~~~\left\vert
\xi\right\vert <x<1\quad.\label{isospin}%
\end{equation}

In the figures are depicted both the vector and axial TDAs, which
explicit expression are given in \cite{Courtoy:2007vy}, for
$m_{\pi}=140$ MeV, $t=-0.5$
GeV$^{2}$ and different values of $\xi$ ranging between $t/(2m_{\pi}%
^{2}-t)<\xi<1$: the process here does not constrain the skewness
variable to be positive. The vector TDA is mainly a function of
$\xi^{2}$ and we have depicted only positive values of $\xi.$ For
the axial TDA, two quite different behaviours are observed
according to the sign of $\xi$. The value we numerically obtain
for $F_{V}\left(  0\right)  $ is in agreement with
\cite{Yao:2006px}, while the one we obtain for $F_{A}\left(
0\right)  $ is twice the expected value \cite{Yao:2006px}.
\begin{figure}[ptb]
\begin{minipage}{8.5cm}
\centering
\includegraphics[width=6.5cm]{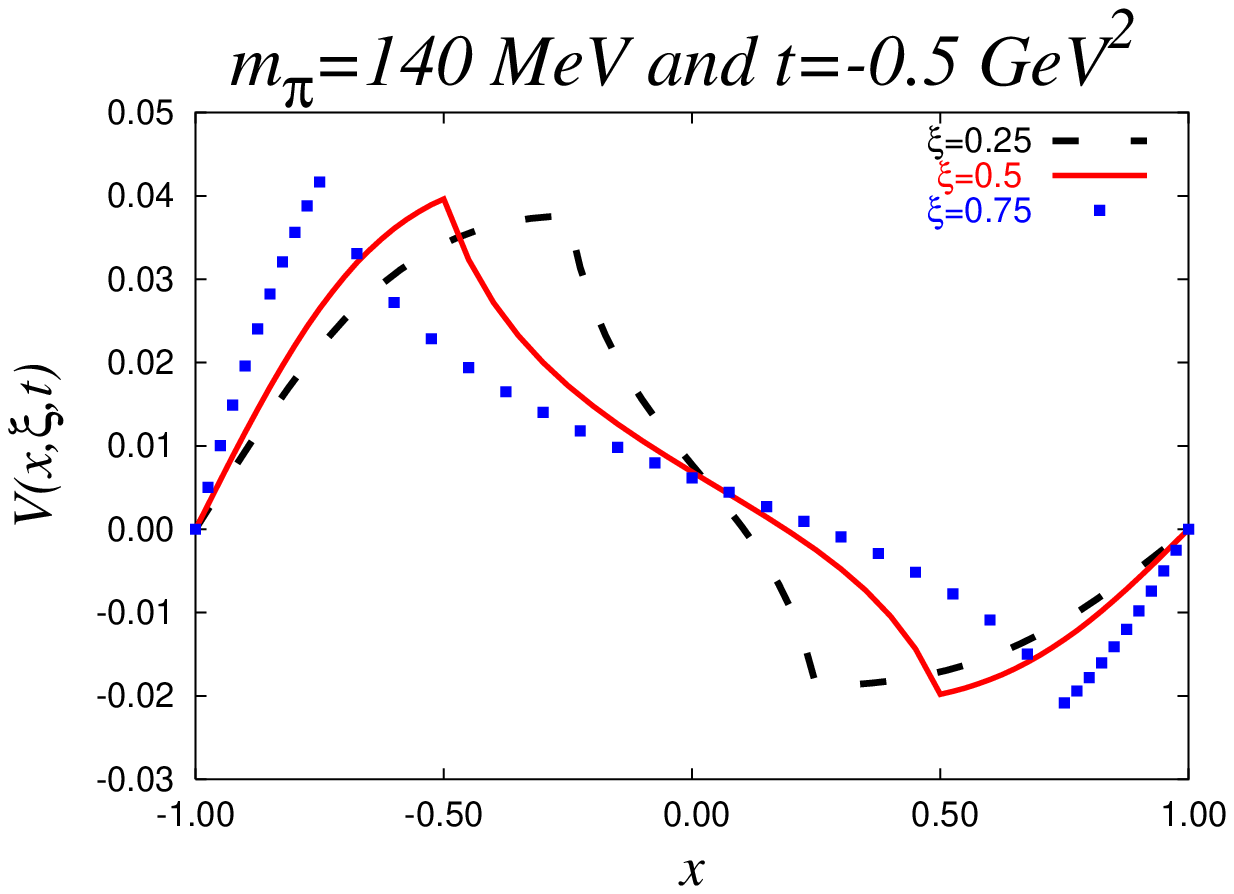}
\end{minipage}
\begin{minipage}{10.5cm}
\centering
\includegraphics[width=6.5cm]{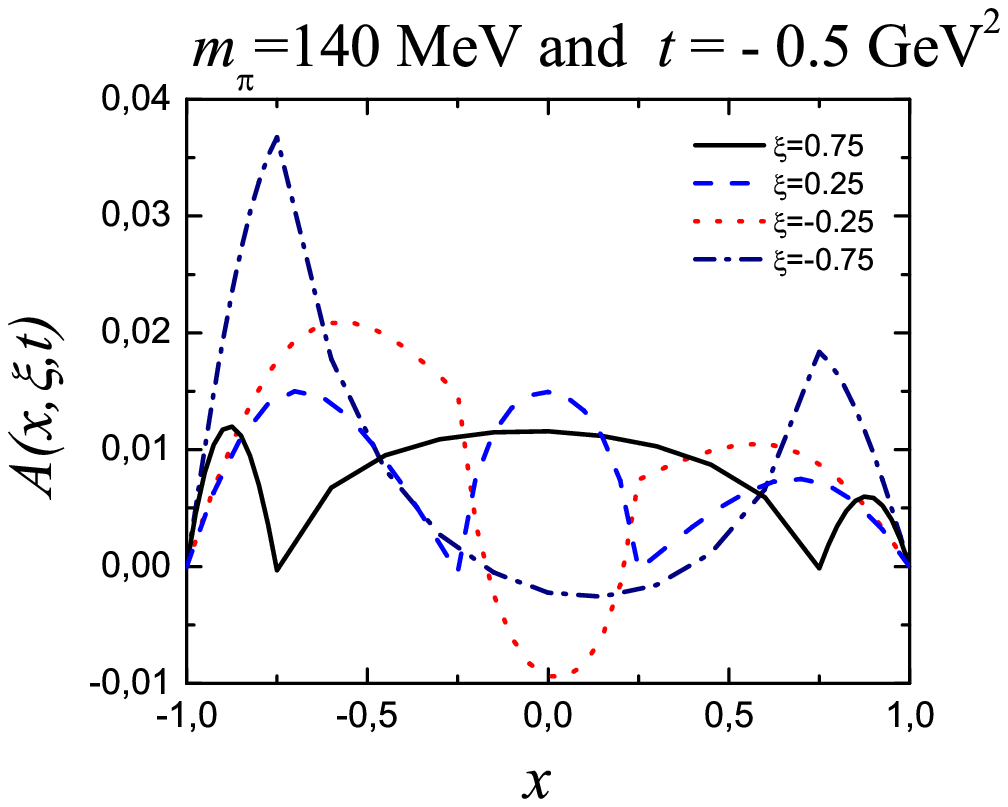}
\end{minipage}
\end{figure}

Previous studies of the pion-photon TDAs have been released
\cite{Tiburzi:2005nj}. Since both these studies parametrize TDAs
by means of double distributions, Ref. \cite{Courtoy:2007vy} is
the first study of the polynomiality property of TDAs. Moreover,
in Ref. \cite{Courtoy:2007vy}, the support, sum rules and
polynomiality expansion are results (and not inputs) of the
calculation. The study of TDAs should lead to interesting
estimates of cross-sections for exclusive meson pair production in
$\gamma\gamma^{\ast}$ scattering \cite{Pire:2004ie}. In
particular, a deeper study of the pion pole contribution should
allow us to give a cross-section estimate for the $\pi\pi$ pair
case.

\end{document}